\documentclass[journal]{IEEEtran}
\usepackage{amsmath,amsfonts}
\usepackage{algorithmic}
\usepackage{algorithm}
\usepackage{array}
\usepackage[caption=false,font=normalsize,labelfont=sf,textfont=sf]{subfig}
\usepackage{textcomp}
\usepackage{stfloats}
\usepackage{url}
\usepackage{verbatim}
\usepackage{graphicx}
\usepackage{cite}
\usepackage{siunitx}
\begin{document}

\title{2.5D co-packaged optical I/O chipsets on a SiON/Si interposer for 4 $\times$ 100G optical interconnection}

\author{Daibao Hou\dag, Yuntian Yao\dag, Xiaotian Cheng, Shuning Ding, Qiyou Wu, Yonghong Hu, Wei Pan, Chao Huang, Huihui Zhu, Yongzhen Huang, ~\IEEEmembership{Fellow, ~OSA,} Chenhui Li, and Chaoyuan Jin*

\thanks{Daibao Hou, Yuntian Yao, Xiaotian Cheng, Shuning Ding, Qiyou Wu, Huihui Zhu, Chenhui Li, and Chaoyuan Jin are with the State Key Laboratory of Silicon and Advanced Semiconductor Materials \& College of Information Science and Electronic Engineering, Zhejiang University, Hangzhou, 310027, China, and also with the ZJU-Hangzhou Global Scientific and Technological Innovation Center, Zhejiang University, Hangzhou 311200, China (e-mail: 12131038@zju.edu.cn; 22431123@zju.edu.cn; chengxt@zju.edu.cn; dingshn@zju.edu.cn; 22431173@zju.edu.cn; huihui001@zju.edu.cn; chenhui.li@zju.edu.cn; jincy@zju.edu.cn).}

\thanks{Yonghong Hu, Wei Pan, and Chao Huang are with the Zhejiang Laychip Optoelectronics Technology Co., Ltd, Taizhou 317500, China (e-mail: brent.hu@laychip.com; wei.pan@laychip.com; chao.huang@laychip.com).}

\thanks{Yongzhen Huang is with the Institute of Semiconductors, Chinese Academy of Sciences, Beijing 100083, China (e-mail: yzhuang@semi.ac.cn).}

\thanks{Chaoyuan Jin is also with the College of Integrated Circuits, Zhejiang University, Hangzhou 311200, China (e-mail: jincy@zju.edu.cn). }

\thanks{Daibao Hou and Yuntian Yao contributed equally to this work. Chaoyuan Jin is the corresponding author.}
}

\markboth{}%
{}

\IEEEpubid{}

\maketitle

\begin{abstract}
Optical I/O technologies have emerged as a potential industrial solution for high-performance data interconnection in AI/ML computing acceleration. While optical I/Os are deployed at the edge of computational chips by co-packaged optics (CPO), flexible and high-performance integration architectures need to be explored to address system-level challenges. In this work, we present and experimentally demonstrate a SiON/Si-based optical interposer that integrates high-bandwidth and energy-efficient optical I/O chipsets. High-performance photonic and electronic components are co-packaged on the interposer, leading to low-loss, signal-integrity-friendly, and thermally efficient characteristics. The optical interposer incorporates low-loss SiON photonic circuits to realize scalable waveguide routing and wavelength-division multiplexing (WDM) with polarization-insensitive operation and high fabrication tolerance, while supporting flip-chip integration with InP-based active devices, including electro-absorption modulated lasers (EMLs) and photodetectors (PDs). Based on this architecture, a 400-Gb/s single-fiber optical transceiver is implemented and experimentally evaluated. Clear eye diagrams and high receiver sensitivity demonstrate reliable high-speed data transmission, which offers scalable, high-bandwidth optical I/Os in future high-performance computational clusters. 
\end{abstract}


\section{Introduction}
\IEEEPARstart{T}{he} rapid growth of artificial intelligence (AI) and machine learning (ML) workloads has imposed unprecedented demands on the bandwidth and energy-efficiency of optical interconnection \cite{wade2020teraphy, tan2023co, lee2022beyond, mahajan2021co}. High-capacity optical I/Os have been proposed as key enabling technologies for optical interconnects between state-of-the-art computational chips. At electro-optic (E/O) interfaces, CMOS-compatible silicon photonics offers a feasible and innovative platform for the convergence of photonic and electronic integrated circuits towards scalable optical I/Os for high-performance computational systems \cite{shi2022silicon}. Very recently, a few demonstrations of high-speed optical I/Os co-packaged with ASIC chips have been reported, whereas optical transceivers are implemented through multi-channel silicon photonics chips, with external optical power provided by off-chip high-power lasers \cite{wade2020teraphy, kannan2024high}. However, the external deployment of laser sources usually requires additional fibers or lenses, which brings in undesired coupling losses, extra packaging volume, and cost. The high-power input causes occasional damage to the facets of the waveguide due to two-photon absorption \cite{liu2007time}. Therefore, heterogeneously integrated \cite{fathololoumi2022highly, piels2023800, takeda2024iii, liang2021recent} or hybrid integrated \cite{gourikutty2024compact, levy20234, bian20233d, marinins2022wafer} lasers on silicon have been proposed for power efficient optical I/Os \cite{fathololoumi2022highly, gourikutty2024compact}. 

On-chip light sources naturally open an alternative technique route towards the integration of EMLs on silicon \cite{ostrovskis2025heterogeneously, hiraki2023over, levy20234}, which offer low power consumption, small footprint, and high modulation bandwidth. Recent advance in native-substrate InP EML with modulation bandwidth exceeding 110 GHz highlight their potentials for high-speed on-chip optical signal generation \cite{chen2025540gbps}. Integrating EMLs together with InP-based PDs onto silicon represents a compelling solution to overcome the limitations of silicon photonics while satisfying the scalability and high edge bandwidth requirements for optical I/Os.

While the above technique route integrates all active components, e.g. lasers, modulators, and PDs, onto the same silicon platform, there leaves a free option for the selection of low-loss materials for integrable passive components \cite{xiang2022silicon, theurer2020flip}. Low-index-contrast photonic circuits based on silicon nitride (SiN) or silicon oxynitride (SiON) relax the constraints on the spot size of the guided mode, and enable low propagation loss and efficient optical coupling with external devices \cite{worhoff2015triplex}. Very recently, active components integrated with low-loss SiN passive waveguides have been demonstrated \cite{bian20233d, norberg2025silicon, xiang2021high}. 

In this work, we propose and experimentally demonstrate on-chip EMLs based optical I/Os using 2.5D co-packaged chipsets, which are fully integrated on SiON/Si optical interposers for 4 $\times$ 100G optical interconnection. Active optical components, including InP EMLs and PDs, and electronic chips, including drivers and transimpedance amplifiers (TIAs), are flip-chip integrated onto the SiON/Si optical interposers. SiON layer serves as a low-loss passive photonic layer that incorporates coarse wavelength-division multiplexing (CWDM) functionalities, realized by arrayed waveguide gratings (AWGs) with high fabrication tolerance for multi-channel lasers, together with low-loss single-mode fiber couplers exhibiting polarization-insensitive characteristics. By combining CWDM photonic circuits with chiplet-based integration, the proposed compact optical I/O architecture provides a scalable and high-performance solution for next-generation high-capacity E/O interfaces.

\section{Architecture of optical I/Os for computational chips}

\begin{figure}[htbp]
\centering\includegraphics[width=8cm]{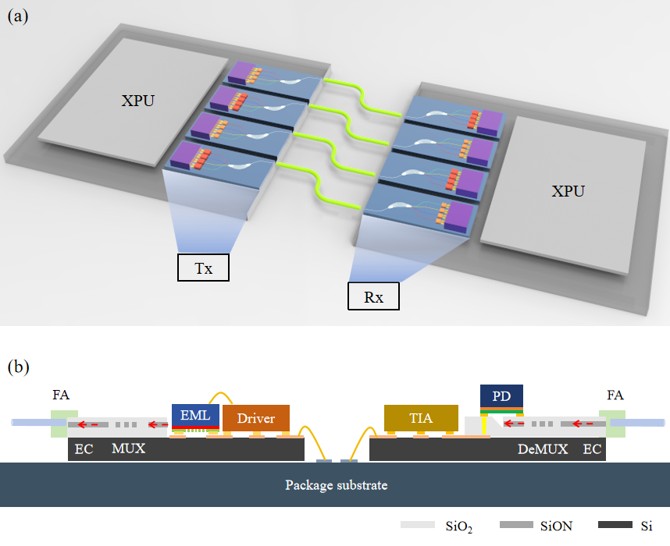}
\caption{Schematic illustration of the Optical I/O architecture based on TX/RX optical engines. (a) Data transmission between computational chips by SiON/Si optical engines; (b) Cross-sectional schematic photograph of co-packaged SiON/Si optical engines.}
\end{figure}

As high-performance computational systems are continuously scaling toward large-scale superpods and superclusters with higher utilization of hardware resources, the bandwidth and energy efficiency of short-reach interconnects have become a critical bottleneck for chip-to-chip data communications. Optical I/Os co-packaged in close proximity to computational chips e.g. XPUs to minimize signal losses have emerged as a potential industrial solution to enable high-bandwidth, low-latency, and energy-efficient communications between electrical chips \cite{fathololoumi2022highly, raghunathan2022scip}. Fig.~1(a) illustrates a schematic architecture for chip-to-chip communication enabled by optical I/Os based on TX/RX optical engines, where bandwidth-limited electrical signals are converted into high-speed optical signals with low propagation loss, thereby extending the escape bandwidth of computational chips and supporting excessive data transmission across large-scale computational fabrics. In practical computational systems, the deployment of optical engines is subject to a combination of stringent system-level constraints. Around the computational chips, the limited perimeter at the chip edge imposes strict limitations on the size and the number of channels of I/Os, driving optical I/Os to deliver ever-higher aggregate bandwidth within a tightly confined space. Meanwhile, electrical data streams must traverse short electrical traces among computational chips, driver circuits, and photonic transmitters before the E/O conversion. Over such short distances, high-frequency electrical signals are highly susceptible to parasitic loss, impedance discontinuities, and signal distortion, making the electrical signal integrity (ESI) a critical design concern. These challenges are further exacerbated by the substantial heat generated by power-hungry computational chips, which places stringent thermal management requirements on co-located optical I/O modules to ensure stable and reliable operation. Therefore, these tightly coupled constraints pose significant engineering challenges and technique demands at the system level for the implementation of optical I/Os. 

In our architectures presented in Fig.~1(b), the electronic chips (drivers/TIAs) and the photonic chips (EMLs/PDs) are all flip-chip bonded onto layered SiON/SiO$_2$/Si structures to guarantee the shortest electrical connections. The drivers convert the external differential electrical signal to single-ended high-speed electrical signal, which is then applied to EMLs and converted into high-speed optical signals. ESI is ensured by placing drivers in close proximity to EMLs, minimizing the path length between the drivers and the modulators to suppress parasitic losses and signal degradation at high frequencies. The dielectric environment provided by the SiON/SiO$_2$ passive layers offers low parasitic loss and excellent electrical isolation, which further benefits the ESI of short-reach electrical links. All drivers/TIA and EML chips directly adhere to the silicon substrate through thin metal contacts and flip-chip technologies to maintain high thermal conductivity for heat dissipation. This configuration alleviates thermal constraints associated with densely integrated optical I/Os around computational chips and supports stable operation under high data-rates. 

In aspects of photonic circuits, apart from the EMLs and PDs being flip-chip bonded, wavelength-division multiplexer (MUX) and demultiplexer (DeMUX) are monolithically integrated in SiON/SiO$_2$ layers based on CWDM protocols that enable high bandwidth density with compact integration. By leveraging these on-chip WDM photonic circuits, the bidirectional optical transmitter and receiver support parallel data transmission over shared optical channels, thereby realizing high-bandwidth optical links for high-speed chip-to-chip communication. The overall energy efficiency of such co-packaged architecture is strongly influenced by losses in the passive photonic integrated circuits. Optical signals traverse passive components including MUX/DeMUX, routing waveguides, and fiber interfaces on the SiON/Si interposer, where accumulated insertion loss directly tightens the optical power budget of transmitters. Consequently, the adoption of low-loss and low-index passive photonic components becomes essential for realizing the energy-efficient optical I/Os for chip-to-chip interconnections. After taking all of these factors into account, these co-packaged chipsets on the SiON/Si interposer provide an attractive solution to enable efficient optoelectronic interface with high performance, thereby forming a solid foundation for high-capacity optical I/O architectures.

\section{SiON/Si optical interposer with Passive Photonic Integration}
As dielectric materials, SiN/SiON exhibit favorable electrical and optical properties, making them particularly attractive for densely integrated electronic–photonic systems. SiN/SiON has excellent insulation characteristics, including high resistivity and low dielectric loss, which effectively suppress parasitic coupling and leakage currents. These properties are beneficial for maintaining high ESI in short-reach and high-speed interconnects. In addition, SiN/SiON materials have recently been adopted in silicon photonics foundries as low-index materials to overcome the physical limitation of Si waveguides in terms of low-loss optical propagation and high coupling efficiency, which further enhance the design capability and the production yield of heterogeneous photonic integration on the silicon photonics platform \cite{wilmart2019versatile, piels2014low}. In this section, we first introduce the advantages of SiON/Si optical interposers, especially for the benefits brought in by the low refractive index. Subsequently, we focus on the practical performance of key passive components integrated on SiON/Si optical interposers, including fiber couplers for SMF and AWGs for WDM. 

\subsection{Advantages of SiON/Si passive photonic integration}
In terms of intrinsic optical properties, SiN/SiON has a broad transparent optical window from the ultraviolet to the mid-infrared, enabling extensive applications in optical communication, sensing, and spectroscopy \cite{worhoff2008silicon, li2013vertical, frigg2019low}. Besides, it has an order of magnitude lower Kerr nonlinear coefficient than silicon and negligible two-photon absorption effects, providing unique advantages in high power handling and nonlinear optics research \cite{baets2016silicon, frigg2019low, gardes2022review}. The thermo-optic coefficient of SiN/SiON is also lower than silicon. The characteristics of temperature insensitivity makes them stable choices in high-temperature environments \cite{gardes2022review}. 

Compared with silicon waveguides, SiON/SiO$_2$ waveguides feature a substantially reduced refractive index contrast between the core and cladding layers, which leads to distinct advantages for passive photonic circuits targeting optical I/O applications. While a high refractive index contrast enables strong optical confinement and ultra-compact photonic circuits, it also amplifies sensitivity to sidewall roughness and fabrication imperfections. According to waveguide scattering theory, propagation loss induced by sidewall roughness scales with \((n_1^2 - n_2^2)^2\) \cite{yin2012fabrication, baets2016silicon}, making high-index-contrast silicon waveguides more susceptible to scattering loss. In contrast, the reduced refractive index contrast of SiON/SiO$_2$ waveguides effectively suppresses scattering loss, enabling low-loss optical routing particularly in the O-band, which is highly desirable for optical I/O power budgeting. The reduced optical confinement in SiON/SiO$_2$ waveguides further facilitates flexible mode conversion by simple geometric design. Compared with tightly confined silicon waveguides with sub-micrometer sizes, the cross-section of SiON/SiO$_2$ waveguides significantly relaxes mode matching and alignment tolerance when interfacing with micron-level optical modes, which enable lower coupling loss at optical input and output interfaces. In addition, the adjustable refractive index of SiON, achieved by controlling the oxygen content during deposition \cite{worhoff2007design}, provides extra degrees of freedom to optimize mode profiles without complex mode converters. From a system-level perspective, the moderate refractive index contrast of optical waveguides offers a flexible design space for polarization-insensitive passive photonic components \cite{baets2016silicon}. Dimensional variations introduced during fabrication lead to relatively small effective index changes in SiON/SiO$_2$ devices, resulting in improved processing tolerance and enhanced wavelength stability for interferometric and wavelength-selective components. These attributes make the SiON/Si material system particularly well suited as a low-loss, fabrication-tolerant passive optical interposer for scalable optical I/O architectures.

\subsection{High performance of SiON components}
AWG is a core building block of integrated photonics for WDM, enabling multi-channel optical signals at different wavelengths to be (de)multiplexed efficiently on single devices. AWGs have been widely demonstrated in WDM transceivers, due to their compact size, scalable channel counts, and compatibility with various photonics platforms \cite{hu2020si3n4, doi2020receiver, andrianopoulos2024integrated}. The performance of AWG-based MUX/DeMUX, including insertion loss, spectral shape and central wavelength stability, significantly affects the overall system performance. 

Fig.~2(a) illustrates a typical Gaussian transmission spectrum of AWG-based MUX for CWDM transmitters. The AWG-based MUX is designed on single-mode SiON/SiO$_2$ waveguides to reduce signal distortion caused by mode dispersion and coupling loss with SMFs, and the insertion loss is around -2.25 dB per channel. However, for DeMUX devices on the receiver side, a flat-top transmission spectrum is desired for the wavelength shift from transmitters. The flat-top spectrum can be obtained by convolving a Gaussian field distribution with the hump-shaped field distribution, which is usually formed by using multimode waveguides at the input or output of AWG \cite{amersfoort1996passband, zheng2020design}. Fig.~2(b) reveals a flat-top passband of around 13 nm with an insertion loss of approximately -1.25 dB per channel, while the crosstalk between adjacent channels remains below -20 dB. Since the multimode waveguide involves more optical modes, this AWG-based DeMUX exhibits lower insertion loss than the MUX \cite{doi2020receiver}. 

In addition to the high performance of MUX/DeMUX devices, the central wavelength drift of MUX/DeMUX is a critical issue in practical applications of WDM technologies. The variation of environment temperatures and the deviation of device sizes cause wavelength shifts compared to the design. The SiON/Si optical interposer exhibits high tolerance for both factors. On one hand, SiN/SiON-based materials have a lower thermo-optic coefficient ($\sim$10$^{-5}$) compared to Si \cite{gardes2022review}. There is a smaller wavelength shift during the temperature change. On the other hand, because of the smaller change in refractive index when the size of device deviates from the design, SiON MUX/DeMUX tends to have a small central wavelength shift. According to experience, nano-scale variation in dimension due to processing errors results in a central wavelength drift of around 1 nm for the SOI photonics platform \cite{baets2016silicon, bucio2016material, ng2015exploring, tan2018nonlinear}. As a result, for AWGs integrated on SiON/Si interposer, the average central wavelength shift of 240 channels is about 1.2 nm for both MUX and DeMUX, with the maximum central wavelength shift of 3 nm, as shown in Fig.~2(c) and Fig.~2(d). 

\begin{figure}[htbp]
  \centering
  \includegraphics[width=8cm]{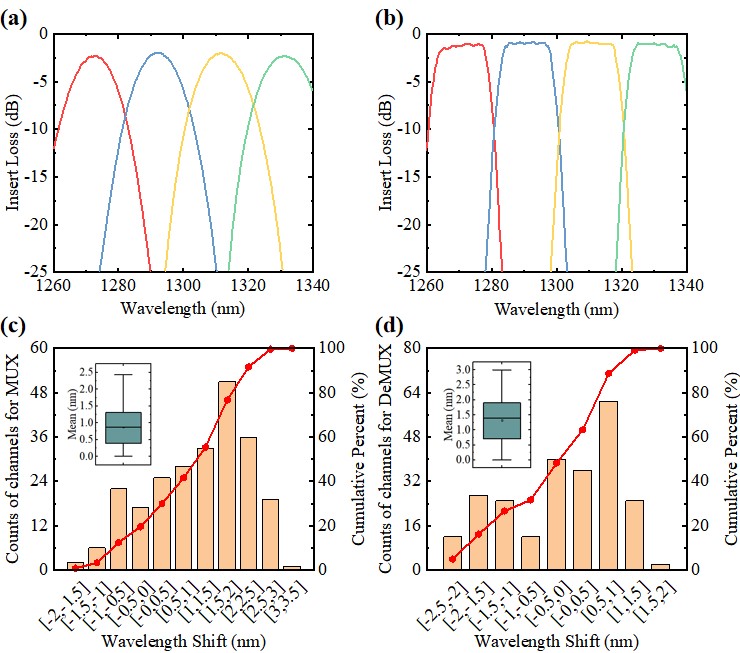}
  \caption{Performance of MUX and DeMUX for O band CWDM. (a) Measured transmission spectrum of MUX; (b) Measured transmission spectrum of DeMUX. (c) The statistics of the central wavelength shift of 240 channels from 60 MUX devices after fabrication; (d) The statistics of central wavelength shift of 240 channels from 60 DeMUX devices after fabrication.}
\end{figure}

As is well known, optical coupling losses are mainly dependent on mode matching. SMFs typically have a mode field diameter(MFD) around 10 microns, which places significant difficulties to achieve high coupling efficiency with silicon waveguides. Although various fiber couplers have been proposed for edge coupling, fine tips and undercut etching are usually required, which are challenging and fragile fabrication tasks. Fortunately, due to the weak optical confinement, the optical mode field can expand flexibly while the size of SiON/SiO$_2$ waveguide narrows down. The optical mode evanescently leaks into the SiO$_2$ caldding layers and is further confined by the SiO$_2$-air interface. The SiON/SiO$_2$ waveguide mode can be flexibly converted into a larger SiO$_2$ waveguide mode for SMF coupling. Fig.~3(a) and Fig.~3(b) show the microscope pictures for the mode field of the fiber coupler with MFD compared to that of the SMF. The insert losses of SiON/SiO$_2$ waveguides with fiber couplers at both sides are tested. As shown in Fig.~3(c), the insert loss is stable at about -1.2 dB in large spectral range from 1260 nm to 1360 nm. Considering the waveguide transmission loss, the fiber coupler exhibits a coupling loss around 0.5 dB/facet. Besides, the similar transmission spectra for TE and TM polarization have proved the polarization-insensitivity of fiber coupler. Fig.~3(d) presented the insert losses of 6 different samples working at 1310 nm, ranging from -0.75 dB to -1.25 dB. We believe the small variation of the insert loss is caused by the misalignment between fiber couplers and SMFs, which also presents the insensitivity of fiber couplers for fabrication. 

\begin{figure}[htbp]
  \centering
  \includegraphics[width=8cm]{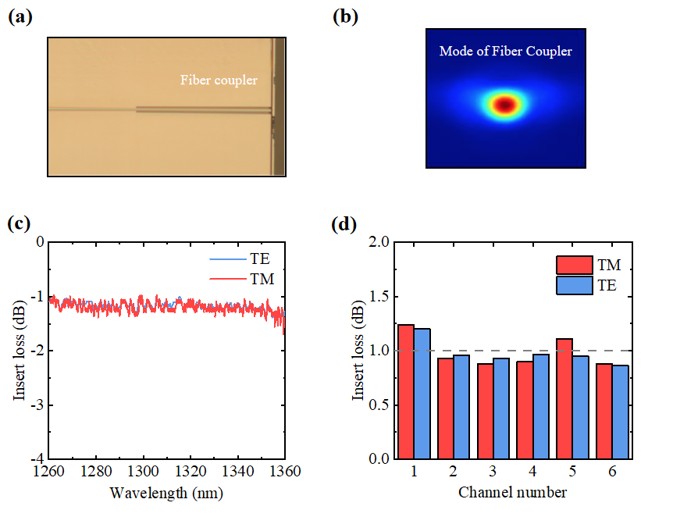}
  \caption{Fiber coupler on the SiON/Si interposer for SMFs. (a) Microscope pictures; (b) Measured optical mode field; (c) Measured transmission spectra under TE and TM polarizations; (d) Measured insert loss of six different samples at 1310 nm.}
\end{figure}

\section{Hybrid Integration of \text{III-V} Devices on SiON/Si interposer}
On the CMOS-compatible silicon photonics platform, various photonic components have been realized for large-scale and high-density integration. However, due to the indirect band gap of silicon \cite{wang2017novel, zhou2023prospects}, monolithic integration of light sources on silicon remains one of the most challenging problems for photonic integration technologies \cite{doerr2015silicon, baets2016silicon, levy20233d, raj202396pj}. Hence, off-chip packaging of III-V lasers has been the standard industrial solution for pluggable optics. When the number of optical channels increases for high-bandwidth transceivers, the output power of III-V lasers needs to increase proportionally for subsequent coupling and splitting. The excessive optical power leads to serious two-photon absorption in silicon waveguides and occasional damage at the coupling facet. Therefor, either the hybrid or heterogeneous integration of III-V lasers directly on silicon chips with temperature insensitive performance is desired for multi-functional photonic integrated circuits \cite{xu2024emerging, jin2006observation, jin2008simple}. 

As a commercially matured method, hybrid bonding technology has been widely applied to the integration of lasers and PDs for various applications\cite{bian20233d, ren2017hybrid, lu2025hybrid}. However, it faces challenges such as effective optical coupling and thermal stability. The former is mainly limited to the mode-size conversion and precise alignment between sub-micron Si waveguides and III-V lasers. Fortunately, SiON/SiO$_2$ waveguide provides an attractive choice for high-efficient edge coupling because of the similar MFD between heterogeneously structured waveguides. In the following, the performance of hybrid-integrated EMLs on the SiON/Si optical interposer is primarily discussed. Statistical data of flip-chip bonded III-V devices, including both lasers and PDs, are presented to prove the stability of device operation.

\begin{figure}[htbp]
\centering\includegraphics[width=8cm]{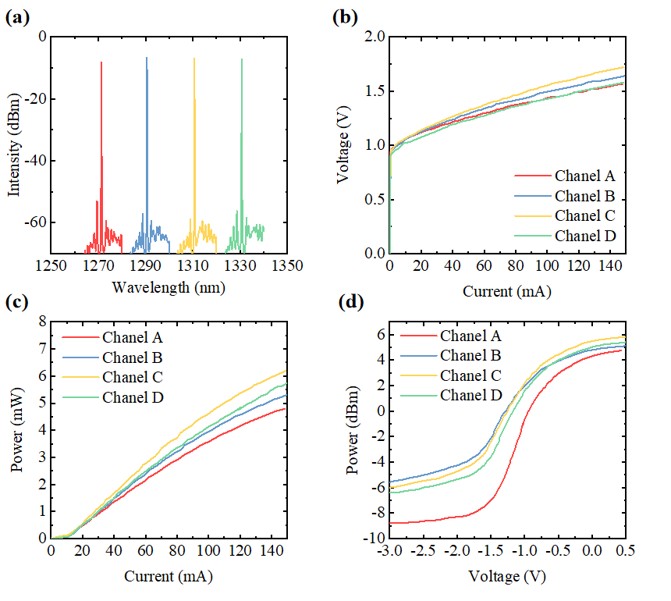}
\caption{The operation performance of four flip-chip bonded EMLs under \SI{50}{\degreeCelsius}. (a) Measured spectra of four CWDM EMLs; (b) Measured I-V curves of four CWDM EMLs; (c) Measured optical power curves at the edge of transmitter chip; (d) Measured modulated optical power curves at the edge of transmitter chip.}
\end{figure}

The optical coupling loss of hybrid integrated lasers is largely dependent on mode mismatching and alignment deviation between heterougenously structured waveguides, along with some contributions from optical scattering from fabrication imperfections and Fresnel reflection at interfaces \cite{bian20233d, bian2021hybrid}. Typically, commercial III-V lasers have an output spot size of 2-3 microns, which is the main reason for high coupling loss when coupling to the mainstream 220nm-thick silicon waveguides. The SiON/SiO$_2$ waveguide provides a micrometer dimension with weak facet reflection simultaneously, which contributes to a better optical coupling with lasers. 

In our work, four EMLs are flip-chip bonded on the transmitter chip. The process of flip-chip bonding through AuSn alloy eutectic welding requires careful optimization for the alignment between chips. The alignment accuracy is usually considered in X, Y and Z directions. The Z-direction alignment between two chips is finely controlled by a mechanical stop structure on active and passive photonic components. The X- and Y-direction alignment is secured by a high-precision bonding machine with an alignment accuracy below 0.5 µm \cite{moscoso2017hybrid}. The alignment accuracy of flip-chip bonding can be improved by considering the freedom of in-plane rotation. 

It is worth noting that although the Fresnel reflection from SiON/SiO$_2$ waveguides is low, a slightly inclined configuration of the waveguide is introduced to reduce the influence. The light emitted by EMLs is coupled into the SiON/SiO$_2$ waveguide, transmitted through the AWG for wavelength routing, and finally coupled into the SMF through the on-chip fiber coupler. Fig.~4(a) presents the emission spectra of EMLs operated at four wavelengths, which are further matched to the center wavelengths of the four channels of AWG. Each laser channel exhibits a side mode suppression ratio (SMSR) exceeding 45 dB. Fig.~4(b) and Fig.~4(c) present the measured I-V and L-I curves under \SI{50}{\degreeCelsius} of the bonded EMLs. According to L-I curves, for the four EMLs, the optical powers are around 5 mW and the threshold currents are about 10 mA. After subtracting the optical loss of AWGs and fiber couplers, the coupling loss between EML and SiON waveguide is estimated in average around -3.5 dB by comparing the optical powers before and after bonding. The modulation of EMLs is measured under a bias current of 100 mA. The fiber power at the edge of the transmitter chip is varied by sweeping the bias voltage from -3 V to 0.5 V, as shown in Fig.~4(d). At the bias voltage of -1 V, the fiber output power is about 1 dBm, and the static extinction ratio is around 7 dB within 1 V$_\textit{pp}$ swing voltage.

\begin{figure}[htbp]
\centering\includegraphics[width=9cm]{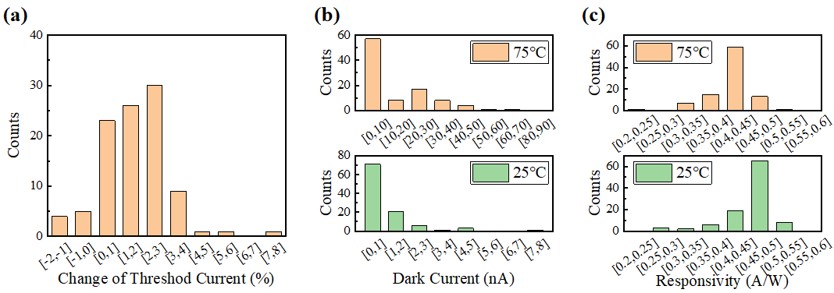}
\caption{Statistics of flip-chip bonded III-V devices. (a)The change of laser threshold current after 2000 hours of operation under 50 mA and \SI{40}{\degreeCelsius}. The dark current (b) and responsiveness (c) of PDs under \SI{25}{\degreeCelsius} and \SI{75}{\degreeCelsius}.}
\end{figure}

Compared with passive silicon photonics devices, III-V devices are more likely to fail after long-time operations. The reliability of III-V devices flip-chip bonded on the SiON/Si interposer is further evaluated, including both lasers and PDs. For 100 integrated III-V lasers, threshold currents are compared before and after 2000 hours of operation under 50 mA and \SI{40}{\degreeCelsius}. The results are shown in Fig.~5(a). Threshold currents of most hybrid integrated lasers change less than 5\%, and the maximum variation is no more than 10\%, which can be attributed to the reliable process of flip-chip bonding and effective heat dissipation through the silicon substrate. Integrated III-V PDs, which are bonded on the receiver chips, were tested under -3V bias voltage. Fig.~5(b) shows the dark current distributions of 80 integrated PDs tested at \SI{25}{\degreeCelsius} and \SI{75}{\degreeCelsius} separately. All PDs present a dark current below 10 nA under \SI{25}{\degreeCelsius}. Even at \SI{75}{\degreeCelsius}, more than 60\% of the devices maintain comparable performance, while the highest dark current among them is no more than 70 nA. Fig.~5(c) shows the responsibility distribution of 80 PDs on receiver chips. The optical power refers to the fiber input power at the edge of receiver chips. The optical signal experiences a total insertion loss less than -2 dB from fiber coupler and AWG-based DeMUX device. The responsivity at \SI{25}{\degreeCelsius} is mostly around 0.475 A/W, and slightly lower at \SI{75}{\degreeCelsius}. All results demonstrate high reliability of integrated III-V devices on silicon by flip-chip bonding for practical applications. 

\section{4 $\times$ 100 Gb/s High-speed Measurements}
The electrical driver and TIA chips were also flip-chip bonded on the SiON/Si optical interposer. The sizes for SiON/Si optical transmitter and receiver chips are 10.21 mm $\times$ 4.74 mm and 9.96 mm $\times$ 3.78 mm, respectively. At the side of transmitter, the differential electrical signal is input to the optical transmitter chip through wire bonding, and the hybrid integrated EMLs are modulated by the single-ended high-speed electrical signals from the driver chips. The optical signals at different wavelengths are combined via the on-chip MUX, and then they are coupled into a SMF of 2km by a low-loss fiber coupler. At the side of receiver, after receiving the modulated optical signals by another fiber coupler, the modulated optical signals are demultiplexed at first into four channels by the flat-top multimode-based AWG and detected by hybrid integrated PDs immediately. The integrated PDs are connected to the nearby TIA so that the responses of PDs are electrically amplified within the shortest distance. Relied on the reduced electrical trace between the optoelectronic chips, the signal distortion and loss can be improved. Both the transmitter chip and the receiver chip present high operation performance under 4 $\times$ 100 Gb/s high-speed measurements. For the transmitter chip, the modulation characteristics and transmitted signal quality are introduced. For the receiver chip, the polarization insensitivity is verified at first. With received high quality optical signals from the transmitter, the receiver sensitivity is further tested. The optical transceiver chips successfully achieve CWDM 400 Gb/s data transmission with low optical loss.

\subsection{Optical signal modulation for transmitter}

\begin{figure}[htbp]
\centering\includegraphics[width=8cm]{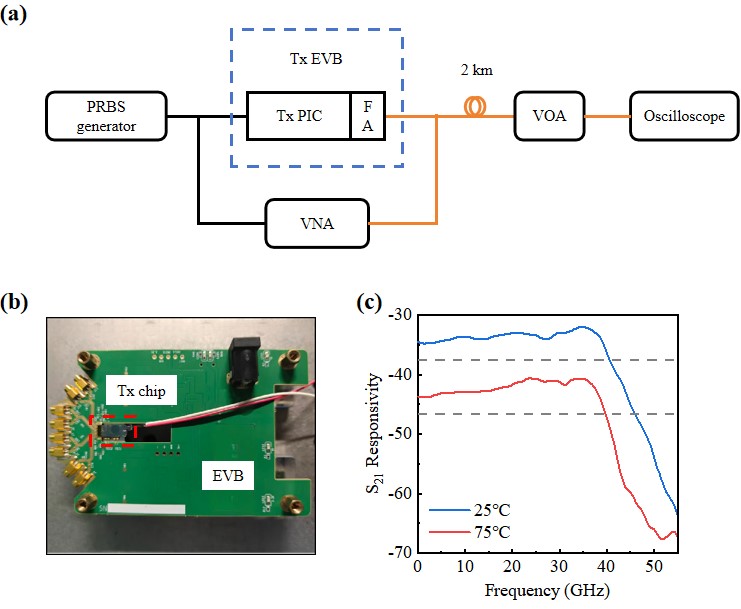}
\caption{ (a) Test link of SiON/Si optical transmitter. (b) Photograph of SiON/Si optical transmitter on EVB; (c) Measured 3dB bandwidth of the SiON/Si optical transmitter.}
\end{figure}

\begin{figure}[htbp]
\centering\includegraphics[width=8cm]{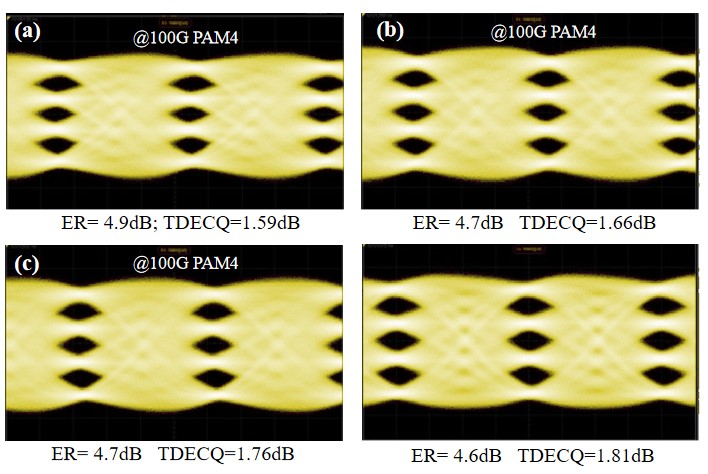}
\caption{ Measured eye diagrams of the SiON/Si optical transmitter.}
\end{figure}

To test the quality of optical signals from the SiON/Si transmitter chip, the test link sets up as shown in Fig.~6(a) and the optical transmitter is packaged on the evaluation boards (EVB) as shown in Fig.~6(b). The electro-optic bandwidth of the transmitter chip is characterized to evaluate the data transmission capability by using vector network analyzer (VNA). Fig.~6(c) presents the frequency response curves from one of the channels. A 3 dB bandwidth around 40 GHz was demonstrated for both \SI{25}{\degreeCelsius} and \SI{75}{\degreeCelsius}. It should be noted that this value refers to the bandwidth of the transmitter containing EMLs, the driver chips, and some high-frequency circuits, providing the capability to achieve more than 100 Gb/s data transmission under PAM4. To further verify the quality of modulated optical signal from this transmitter, the 106.25 Gb/s PRBS31 electrical signal from PRBS generator was sent to the transmitter so that the hybrid integrated EMLs can be modulated. After the modulated optical signal was coupled into SMF from the transmitter, it was attenuated to a suitable power by a variable optical attenuator (VOA), and fed into the high-speed sampling oscilloscope (N1092C) to analyze the time domain characteristics by generated eye diagrams. As shown in Fig.~7(a)--(d), the quality of 106.25 Gb/s PAM4 optical signal generated under \SI{50}{\degreeCelsius} can be evaluated for every channel, and all eye diagrams clearly exhibit open eyes with a maximum TDECQ of 1.81 dB and a dynamics extinction ratio of 4.9 dB.

\subsection{BER testing for polarization-insensitive optical receiver}

\begin{figure}[htbp]
  \centering
  \includegraphics[width=8cm]{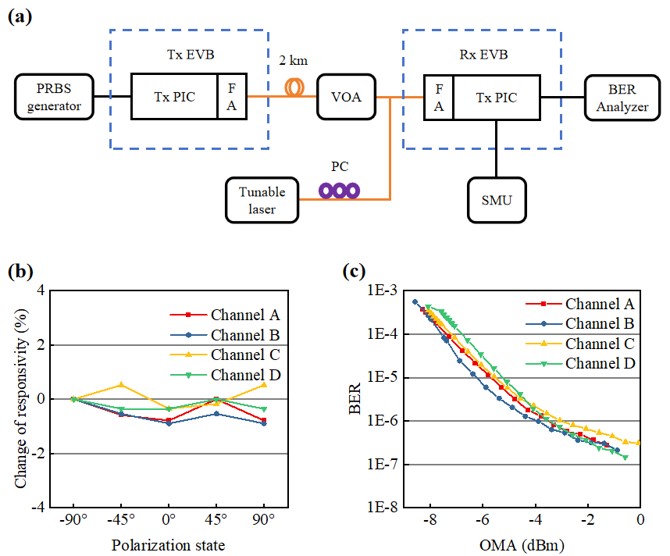}
  \caption{(a) Test link of SiON/Si optical reciever; (b) Test results for the polarization dependent performance; (c) Sensitivity of optical receiver under 4 $\times$ 100 Gb/s PAM4 data transmission.}
\end{figure}

For the receiver chip, polarization sensitivity is a crucial issue, as optical signals with different polarizations cause additional losses and distortions. Polarization-maintaining fibers or polarization controllers are usually required to solve the problem \cite{kim2019polarization, zhao2018broadband,ma2015symmetrical, zhao2024ultracompact}. Here, the polarization insensitivity of the optical receiver chip is demonstrated according to the change in optical signals with different polarization states. The polarization state of input optical signal can be controlled by a polarization controller, and the electrical responses of receiver is directly detected by a SMU voltage source connected with probes. As shown in Fig.~8(b), the change in response from 4 different channels is tested and recorded under 5 different input polarization states. After all responses are normalized to the initial responses, every channel's response exhibits negligible variations, confirming excellent polarization insensitivity for optical receiver chips. 

With the polarization insensitivity of the receiver, the performance of data transmission was further characterized by the bit error rate (BER) test without forward error correction (FEC). After the 106.25 Gb/s PAM4 optical signal emitted from transmitter is transformed into electrical signal and further amplified by the receiver chip, the BER data of every channel can be recorded by a BER analyzer (BA4000). Furthermore, the modulated optical signal is attenuated by the VOA before being input into receiver chip, so that the BA4000 can record the BER data at different levels of input power to test the sensitivity of receiver. The test of BER for every channel is done without the crosstalk of other channels, and the results are shown in Fig.~8(c). An optical modulation amplitude (OMA) sensitivity of approximately -7.25 dBm is achieved at a BER of 2.4 $\times$ 10$^{-4}$ and the BER is close to 1 $\times$ 10$^{-7}$ at an OMA of 0 dBm.

\section{Conclusions}

Integrated in a SiON/Si optical interposer, all-on-chip SiON/Si optical I/O chipsets are realized and subsequently the transmission of the 4 $\times$ 100G PAM4 data is verified. We introduce the main advantages of the SiON/Si-based optical interposer architecture for optical interconnection, including reliable integration of III-V devices and low-loss passive components with polarization insensitivity and high fabrication tolerance. The entire optical link has low optical loss due to the simple link and low-loss optical devices involved. Considering the optical losses attributed to the coupling of laser, MUX/DeMUX and two fiber couplers, the total power budget of the optical transceiver is around 8 dB, which significantly reduces the power requirement for lasers. In fact, with careful design of edge coupling for lasers, the loss can be further reduced. The optical transceiver was verified by 4 $\times$ 100G PAM4 data transmission over 2km of SMF. The SiON/Si optical transmitter presents great capability of E/O conversion with clear eye diagrams, and the SiON optical receiver presents a low per-FEC-BER with polarization-insensitivity.

For practical applications of data transmission, high-bandwidth and energy-efficiency optical interconnection becomes more important than ever before. Photonic integrated circuits comprise all kinds of functional devices are desired, especially for high-performance computational systems. In addition to the loss of signal, the tightened space budget around electronic chips limits the edge bandwidth density. On low-loss SiON/Si optical interposer, lasers can be integrated without additional fibers. The optical loss and bandwidth density can be further improved by careful optimization in the design. In short, the SiON/Si optical interposer architecture has realized all-on-chip integration with low loss, high signal integrity for high-bandwidth and energy-efficiency optoelectronic integration, which may shine a way to the deployment of optical I/Os around computational chips in the future.

\section*{Acknowledgments}
This work is supported in part by the National Innovation Program for Quantum Science and Technology(2023ZD0300800), the National Key Research and Development Program of China (2021YEB2800500), the Zhejiang Natural Science Foundation (LJHSD26F040001), the Zhejiang Province Leading Geese Plan (2024C01105), the Beijing Natural Science Foundation (L248103), and the National Natural Science Foundation of China (61574138, 61974131).


\bibliography{References}
\bibliographystyle{IEEEtran}






\vfill

\end{document}